\documentclass[twocolumn,showpacs,preprintnumbers,prl]{revtex4}
\usepackage{amssymb}

\usepackage{graphicx}

\begin{document}

\title{Short-Term Memory in Orthogonal Neural Networks}
\author{ Olivia L. White$^{*}$, Daniel D. Lee$^{\dag}$, and Haim Sompolinsky$^{*\ddag}$}
\affiliation{$^{*}$Harvard University, Cambridge, MA 02138\\
$^{\dag}$University of Pennsylvania, Philadelphia, PA 19104\\
$^{\ddag}$Racah Institute of Physics and Center for Neural Computation,\\
Hebrew University, Jerusalem 91904, Israel }
\date{\today }
\pacs{89.70.+c, 02.50.-r, 05.20.-y}

\begin{abstract}
We study the ability of linear recurrent networks obeying discrete
time dynamics to store long temporal sequences that are retrievable
from the instantaneous state of the network.  We calculate this
temporal memory capacity for both distributed shift register and
random orthogonal connectivity matrices.  We show that the memory
capacity of these networks scales with system size.
\end{abstract}

\maketitle

The brain holds information in short-term memory for use in
prospective action. It is thought that persistent firing patterns in
cortical networks subserve such working memory and several mechanisms
have been proposed \cite{wang}. In some, a stimulus, such as a spoken
word or a picture activates a pattern of neuronal activity that
persists for several seconds because it corresponds to an isolated
stable fixed point of the dynamics \cite{usher}.  However, working
memory often involves memorization of graded signals, such as stimulus
location in 2D space or the eye's gaze, which are hard to associate
with {\em discrete} attractors.  Recent models propose that short-term
memory is associated with low dimensional {\em continuous manifolds}
of attractors. Both network \cite{seung} and single cell mechanisms
\cite{loewenstein} have been implicated in generating these manifolds.

More recently it was suggested that a {\em generic} recurrent network
can store arbitrary temporal inputs in its {\em transient}
responses, even though these responses do not correspond to
attractors, and that working memory operates by reconstructing input
history from the network's current state \cite{maass, jaeger}. This
proposal has been investigated numerically for some recurrent networks
but its theoretical underpinnings are unexplored. For instance, how
does a network's storage capacity for temporal memory scale with
system size? What network architectures are suitable? How do noise and
structural perturbations affect memory? Such issues have important
implications for general dynamical systems. To what extent can the
history of perturbations on a dissipative system be reconstructed from
its current state?  How does the transient memory of a dynamical
system depend on its number of degrees of freedom and the amount of
noise? In this paper we develop theoretical understanding of the
capacity of linear recurrent networks to store temporal signals.

\begin{figure}[htb]
\centering \includegraphics[width=2.5in]{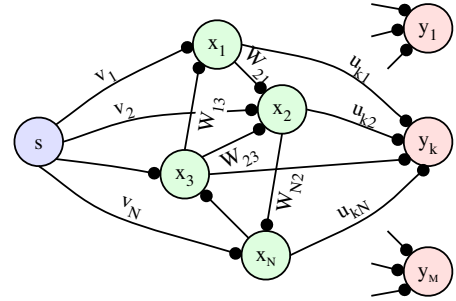}
\caption{Architecture of the short-term memory network. A 
single input unit, $\bf{s}$, feeds into an $N$ unit recurrent network. 
Each of an array of readout units reconstructs the input at a given past 
time.}
\label{fig:model}
\end{figure}

\textbf{Model:} We consider here the discrete time model proposed by
Jaeger \cite{jaeger}.  A time dependent scalar signal, $s(n)$, is
memorized by a linear recurrent network with N neurons obeying the
discrete time dynamics:
\begin{equation}
\mathbf{x}(n)=\mathbf{W}\mathbf{x}(n-1)+\mathbf{v}s(n)+\mathbf{z}(n).
\label{dynam}
\end{equation}
$\mathbf{x}(n)$ gives the network state at time $n$, $\mathbf{v}$ is a
unit norm constant vector of connections from the input source,
$\mathbf{W}$ is the matrix of recurrent connections which need not be
symmetric, and $\mathbf{z}(n)$ is a noise vector. To ensure dynamical
stability, we require $\alpha <1$, where $\alpha$ is the norm squared
of $\mathbf{W}$'s largest eigenvalue. The goal is to extract the
scalar history of the signal, $\{s(m) | m\leq n\}$, from the current
network state $\mathbf{x}(n)$.  This is achieved using a layer of
linear readout neurons, the state of which at time $n$ is given by
$\left\{
y_{k}(n)=\mathbf{u_{k}^{\mathrm{T}}}\mathbf{x}(n),k=0,1,2,...\right\}
$, see Fig.~\ref{fig:model}.  $\mathbf{u}_{k}$ is a constant vector of
output connections from the recurrent network to the $k$-th readout
neuron. It is chosen to minimize mean square deviations
$\langle|y_{k}(n)-s(n-k)|^{2}\rangle_{n}$ so that $y_{k}(n)$ is close
to $s(n-k)$, where $\langle \dots \rangle_{n}$ denotes a time
average. The resultant optimal output weight is
$\mathbf{u_{k}}=\mathbf{C}^{-1}\mathbf{p_{k}}$ where
$\mathbf{C=}\left\langle \mathbf{x}(n) \mathbf{x^{\mathrm{T}}}(n)
\right\rangle_{n}$ is the covariance matrix of $\mathbf{x}$, and
$\mathbf{p_{k}}=\left\langle s(n-k)\mathbf{x}(n) \right\rangle_{n}.$
The ability to embed signals in the network may depend on their
statistics. Here we characterize the signal ensemble by $\langle
s(n)\rangle _{n}=0$ and $\langle s(n)s(n+k)\rangle _{n}=\delta
_{k,0}$.  The noise vectors have zero mean, $\langle \mathbf{z}(n)
\rangle_{n}=0$ and variance $\langle \mathbf{z}_{i}(n)
\mathbf{z}_{j}(n+k)^{T} \rangle_{n}=\epsilon \delta
_{k,0}\delta_{i,j}$.  With the above signal and noise statistics,
$\mathbf{p_{k}}=\mathbf{W}^{k}\mathbf{v}$, and
\begin{equation}
\mathbf{C}=\sum_{k=0}^{\infty}
\mathbf{p_{k}}\mathbf{p_{k}}^{\mathrm{T}}
 +\epsilon \mathbf{C_{n}}
\label{Cgen}
\end{equation}
where the scaled noise covariance is
$\mathbf{C_{n}}=\sum_{k} \mathbf{W}^{k}\mathbf{W}^{kT}$.

\textbf{Memory function:} We define the system's memory function as
the overlap between the past input and its reconstructed value, $m(k)=
\left\langle s(n-k)y_{k}(n) \right\rangle_{n}$ . With the above
statistics,
\begin{equation}
m(k)=\mathbf{p_{k}^{\mathrm{T}}}\mathbf{C}^{-1}\mathbf{p_{k}}.
\end{equation}
$m(k)=1$ corresponds to perfect reconstruction, $y_{k}(n)=s(n-k)$,
whereas $m(k)=0$ indicates no memory of $s(n-k)$. For an arbitrary,
stable connection matrix $\mathbf{W}$,
\begin{equation}
\sum_{k=0}^{\infty }m(k)=\mathrm{Tr}\mathbf{C}^{-1}\sum_{k=0}^{\infty }
\mathbf{p_{k}p_{k}^{\mathrm{T}}}=N-\epsilon \mathrm{Tr}\mathbf{C}^{-1}
\mathbf{C_{n}}.
\label{sumrule}
\end{equation}

This sum rule provides a useful indication of the network's short term
memory.  For zero noise, the area under the memory function is exactly
$N$, implying that all $N$ degrees of freedom are useful for storage.
Storage capacity decreases with strength of noise. System performance
also relates to the $shape$ of $m(k)$.  Since $0\leq m(k)\leq 1$, it
follows that the length of signal that can be exactly reproduced is
also bounded.  To characterize the length of time over which a signal
can be retrieved with reasonable accuracy, we define the temporal
capacity $k_{C}$ as the minimum value of $k$ such that $m(k) <
\frac{1}{2}$.  We focus particularly on the conditions under which the
system's capacity is {\em extensive}, namely $k_{C}\varpropto N$ as
$N \rightarrow \infty$.  For given noise $\epsilon$, we define
$\alpha^{opt}$ as the value of $\alpha$ at which capacity achieves its
maximum, $k_{C}^{opt}$.

\begin{figure}[htb]
\centering\includegraphics[width=3.0in]{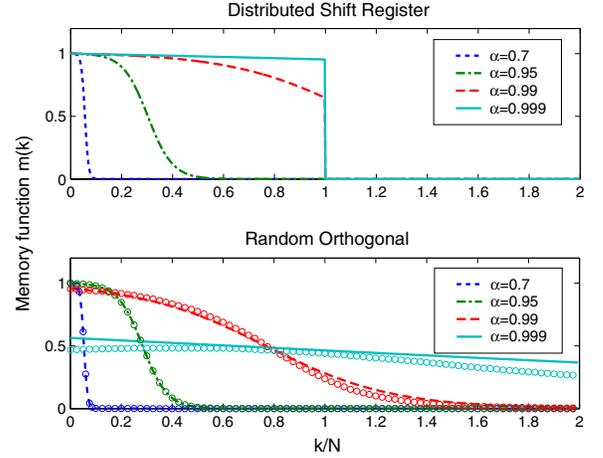}
\caption{Memory capacities of DSR and Orthogonal networks for
  $\epsilon=10^{-4}$ and $N = 400$ at various values of $\alpha$. For
  the Orthogonal network, the circles show simulation results and the
  solid lines show predictions of the Annealed Approximation, which
  begins to break down for $\alpha$ very near 1. While the memory of
  the DSR improves with increasing $\protect\alpha$, the capacity of
  the orthogonal network peaks at $\protect\alpha^{opt}=0.98$. }
\label{fig:examples}
\end{figure}

\textbf{Distributed Shift Register (DSR) Network:} A straightforward
candidate for a short-term memory system is a delay line, or a
{\em shift register} network, which corresponds to a
one-dimensional network with $W_{ij}=\sqrt{\alpha }\delta
_{i,j+1}$, and $v_{i}=\delta _{i,1}$. One drawback of this system is
its extreme sensitivity to removal of a single neuron.  A more robust,
distributed architecture of the shift register operation is a fully
connected network with
\begin{equation}
\mathbf{W}=\sqrt{\alpha }\sum_{k=1}^{N-1} \mathbf{v^{(k+1)} v^{(k)\mathrm{T}}};
 \quad \mathbf{v^{(1)}} \equiv \mathbf{v}
\label{hebb}
\end{equation}
where $\left\{ \mathbf{v^{(k)}}\right\} $ is an arbitrary set of $N$
orthonormal vectors.  Note that $\mathbf{Wv^{(k)}} = \sqrt{\alpha}
\mathbf{v^{(k+1)}}$ for $k \leq N-1$ and $\mathbf{Wv^N}=0$, implying that
$\mathbf{W}^N = 0$
\footnote{Since $\mathbf{W}^N=0$, $|\mathbf{x}|$ converges for any
  finite system size $N$. However, in order for $|\mathbf{x}|$ to not
  grow exponentially with $N$ we still require $\alpha <1$.}.
In this network the covariance matrix is
\begin{equation}
\mathbf{C}=\sum_{k=1}^{N}\left[ \alpha ^{k-1}+\widetilde{\epsilon}(1-\alpha
^{k}) \right]
\mathbf{v^{(k)}}\mathbf{v^{(k) \mathrm{T}}},
\label{Chebb}
\end{equation}
where $\widetilde{\epsilon} = \epsilon/(1-\alpha)$.  The memory function is
given by
\begin{equation}
m(k)=\frac{\alpha^{k}}{\alpha ^{k}+\widetilde{\epsilon}(1-\alpha ^{k+1})},
\;\;k=0,...,N-1
\label{memhebb}
\end{equation}%
and $m(k)=0$ for $k\geq N$.  In zero noise, $\mathbf{x}(n)
=\sum_{k=0}^{N-1}\alpha ^{k/2}s(n-k)\mathbf{v^{(k+1)}}$.  Thus the
network embeds each of the previous $N$ signal values in a distinct
orthogonal direction $\mathbf{v^{(k)}}$ for $k$ up to $N-1$. An
important question in both this and the following models is how the
value of $\alpha$ affects the system performance.  In the absence of
noise the present model retrieves perfectly the most recent $N$ inputs
for all values of $\alpha$, as implied by Eq.~\ref{memhebb}. However, the
required readout weights $\mathbf{u_{k}=}\alpha ^{-k/2}\mathbf{v^{(k+1)}}$
for the retrieval of these memories increase with decreasing $\alpha$, 
limiting the choice of $\alpha$ to values close to 1.

Non-zero noise contributes to $\mathbf{x}(n)$, polluting the signal
but leaving fixed the directions along which temporal signals are
embedded. When $\epsilon>0$, $m(k)<1$ for all $k$.  The capacity
$k_{C}$ is greater than $0$ for all $\epsilon < 1$.  It increases with
decreasing $\epsilon$ and saturates to $k_{C}=N$ at zero noise. For
fixed noise, increasing $\alpha$ increases signal-to-noise ratio and
hence $m(k)$ increases, see Fig.~\ref{fig:examples}. Thus $\alpha
^{opt}=1$ for all values of noise $\epsilon>0$.

\textbf{Random Orthogonal Network:} We next ask whether broader
classes of connection matrices can also store long temporal signals.
A plausible extension of the above model is to a network with
$\mathbf{W}= \sqrt{\alpha}\mathbf{O}$ where $\mathbf{O}$ is an
$N\times N$ orthogonal matrix (i.e., $\mathbf{OO}^{T}=1$) and $\alpha
<1$. Similar to the DSR model, $\mathbf{W}$ performs a rotation
followed by a shrinking with factor $\sqrt{\alpha}$. However
$\mathbf{W}^k \mathbf{v}$ and $\mathbf{W}^{k+1} \mathbf{v}$ are
not necessarily orthogonal, in contrast to the DSR.
Moreover while $\mathbf{W}^{N}=0$ for the DSR, orthogonal $\mathbf{W}$
is full rank. Consequently, inputs from times earlier than $N$ can
interfere with current inputs.  For any choice of $\mathbf{O}$ and
$\mathbf{v}$ the covariance matrix is
\begin{equation}
\mathbf{C}=\sum_{k=0}^{\infty }\alpha ^{k}\mathbf{O}^{k}\mathbf{vv}^{T}
\mathbf{O}^{-k}+\widetilde{\epsilon }\mathbf{I},\;\;
\widetilde{\epsilon} \equiv \epsilon /\left(1-\alpha \right).
\end{equation}
However the system behavior can depend upon the particular
$\mathbf{O}$ and $\mathbf{v}$. We therefore consider $\mathbf{O}$
drawn from the Gaussian Orthogonal Ensemble (GOE) and input
connections $\mathbf{v}$ from a Gaussian distribution with
$\mathbf{v}^{T}\mathbf{v}=1$. We evaluate $m(k)=\langle
\mathbf{p_{k}^{\mathrm{T}}C^{\mathrm{-1}}p_{k}} \rangle$ where the
average is over these ensembles and captures the typical
behavior for $N$ large.  Exact analytical evaluation of $m(k)$ is
complex and requires accounting for statistical correlations between
powers of random orthogonal matrices. In the following we solve the
problem under the ``Annealed Approximation'' (AA), in which
$\mathbf{W}$ and $\mathbf{v}$ are not quenched in time but drawn
randomly at each time step. Under this approximation,
\begin{equation}
m(k)=\frac{\alpha ^{k}q}{1+\alpha ^{k}q}
\label{annealed}
\end{equation}
where $q$ satisfies:
\begin{equation}
1=N^{-1}\sum_{k=0}^{\infty }\frac{\alpha ^{k}q}{1+\alpha ^{k}q}+
 \widetilde{\epsilon}q.
\label{sumannealed}
\end{equation}
To see this, first note that in the annealed scenario $\mathbf{p}_{k} = 
\mathbf{v^{(k)}}$ where $\left\{ \mathbf{v^{(k)}}\right\} $ is an infinite
set of independent random normalized vectors.  Hence,
\begin{equation}
m(k)=\left\langle \alpha ^{k}\mathbf{v^{(k)\mathrm{T}}}\left[ \mathbf{I}+%
\mathbf{C_{0}^{\mathrm{-1}}}\alpha ^{k}\mathbf{v^{(k)}v^{(k)\mathrm{T}}}%
\right] ^{-1}\mathbf{C_{0}^{\mathrm{-1}}v^{(k)}}\right\rangle
\end{equation}%
where $\mathbf{C_{0}}\equiv \mathbf{C}-\alpha^{k} \mathbf{v^{(k)}v^{(k)%
\mathrm{T}}}$ is independent of the random variable $\mathbf{v^{(k)}}$.
Expanding in powers of $\alpha^{k}$ and averaging over $\mathbf{v^{(k)}}$
yields Eq.~(\ref{annealed}) for the memory function where
\begin{equation}
q=\left\langle \mathbf{v^{(k)\mathrm{T}}C_{0}^{\mathrm{-1}}v^{(k)}}%
\right\rangle =\frac{1}{N}\left\langle \mathrm{Tr}\mathbf{C_{0}^{\mathrm{-1}}%
}\right\rangle =\frac{1}{N}\left\langle \mathrm{Tr}\mathbf{C^{\mathrm{-1}}}%
\right\rangle .
\label{q}
\end{equation}
The last equality holds for large $N$ because $\left\langle
\mathrm{Tr}\mathbf{C^{\mathrm{-1}}}\right\rangle -\left\langle
\mathrm{Tr}\mathbf{C_{0}^{\mathrm{-1}}}\right\rangle \thicksim
\alpha^{k} \left\langle
\mathbf{v^{(k)\mathrm{T}}C_{0}^{\mathrm{-2}}v^{(k)}}\right\rangle$,
which is only of order 1 due to the normalization of
$\mathbf{v^{(k)}}$.  Eq.~(\ref{sumannealed}) for $q$ is obtained
from the sum rule Eq.~(\ref{sumrule}) by substituting
$\mathbf{C_{n}}=(1-\alpha )^{-1}\mathbf{I}$. Though the annealed
approximation neglects quenched correlations, it agrees surprisingly
well with the numerical solution of the quenched system for all
$\alpha$ values except for $\alpha \rightarrow 1$, seen in the examples of
Fig.~\ref{fig:examples} for $\epsilon = 10^{-4}$.

\begin{figure}[htb]
\centering \includegraphics[width=3.0in]{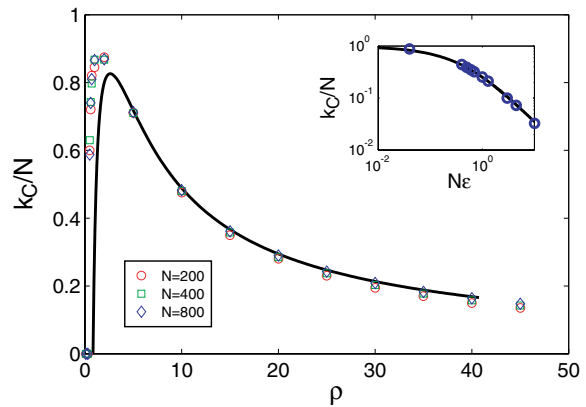}
\caption{Capacity per neuron as a function of $\rho=N(1-
\alpha)$, at $\bar{\epsilon}=0.04$, showing non-monotonic
dependence.  Points from simulations on differently sized systems fall on
essentially the same curve, confirming that in this regime capacity is
extensive. The inset shows the dependence of $k_C/N$ on scaled noise $\bar{
\epsilon}$ for $\rho^{opt}$ both in the AA and from
simulation.}
\label{fig:extensive}
\end{figure}

To analyze the network's behavior, we first consider the limits 
$(1-\alpha)/N$, $\epsilon /N\rightarrow0$ for $N\rightarrow \infty$. 
In this case, the sum of $m(k)$ is finite so Eq.~(\ref {sumannealed}) 
gives $q=1/\tilde{\varepsilon},$ and thus Eq.~(\ref{annealed}) reduces to
\begin{equation}
m(k)=\frac{\alpha ^{k}}{\alpha ^{k}+\tilde{\epsilon}},\;\;k=0,1,2,...
\label{memfinite}
\end{equation}
Capacity is nonzero for $\tilde{\epsilon}<1,$ in which case
$k_{C}=\log (\tilde{\epsilon})/\log (\alpha)$. Here
$\alpha^{opt}(\epsilon)$ is less than unity and decreases with
increasing noise because it results from a balance between signal
suppression on one hand and amplification of error input $\epsilon$ by
the factor $(1-\alpha )^{-1}$ on the other.  For small $\epsilon$,
maximizing $k_{C}$ with respect to $\alpha$ yields $\alpha
^{opt}\approx 1-\epsilon e$ and $k_{C}(\alpha ^{opt})\approx
1/\epsilon e$.  For $\epsilon \rightarrow 1$, $\alpha
^{opt}\rightarrow 0$ and $k_{C}\rightarrow
0$. Equation~(\ref{memfinite}) is exact in the limit of large $N$ if
$\epsilon$ and $\alpha$ are kept fixed.

In order to yield extensive capacity, $\epsilon$ and $1-\alpha$ must
decrease inversely with $N$, so that $1-\alpha =\rho /N$ and $\epsilon
= \bar{\epsilon}/N$ with $\rho$ and $\bar{\epsilon}$ finite. To see
that in this case $k_{C}$ scales with system size $N$, we write
$q=\exp (\rho \mu )$.  Capacity $k_C = \mu N$ is extensive for $\mu =
O(1)$ and $k_{C}=0$ for $\mu <0$, see Eq.~(\ref{annealed}). The sum
rule Eq.~(\ref{sumannealed}) determines the value of $\mu$. In the
present regime the sum can be approximated by an integral, yielding
\begin{equation}
\rho =\log (1+\exp (\rho \mu ))+\bar{\epsilon}\exp (\rho \mu ).
\label{sumextensive}
\end{equation}
Solving for $\mu$ yields a nonmonotic function of $\rho$ (see
Fig.~\ref{fig:extensive}) which attains its maximum at 
$\rho^{opt}$.
For $\rho < \rho^{opt}$, $\mu$ decreases
with decreasing $\rho$ and $\mu<0$ for $\rho$ smaller than the critical
value $\rho_{-}=\log (2)+\bar{\epsilon}$.  
For large noise, i.e., $\bar{\epsilon} \gg 1$, $\rho^{opt}$ increases
with noise level as $\rho^{opt} \approx e \bar{\epsilon}$, as in the 
finite capacity limit.  However, at low noise levels $\rho^{opt}$
does not approach 0 (as predicted by the finite capacity limit)
but increases with decreasing $\bar{\epsilon}$ as
$\rho^{opt}\approx \frac{1}{2}\log(1/\bar{\epsilon})$. 
This is because for $\bar{\epsilon}$ sufficiently small and $\alpha$ 
sufficiently close to 1, strong long-time interference
prevents faithful reconstruction even of recent inputs.  Therefore,
$\alpha^{opt} < 1$ even in the $\bar{\epsilon} \rightarrow 0$ limit.
Additionally, choosing $\rho$ close to
$\rho^{opt}$  reduces retrieval error for
values of $k < k_C$.  This behavior is demonstrated in
Fig.~\ref{fig:examples}. As in the DSR model, the choice
of $\rho$ should be bounded not only by capacity limits but also by
the magnitudes of the output weights which increase with $\rho$
roughly as $||\mathbf{u_k}|| \approx \alpha^{-k/2}=\exp(\rho k/N)$ 
at zero noise.

\textbf{The Annealed Approximation:} An interesting issue is the range
of validity of the annealed approximation. As indicated above, finite
capacity results should be exact in the large $N$ limit with fixed
noise and suppression coefficient.  In this limit, at any given time
only a small number of directions in $\mathbf{x}$ space contribute to
the current state and since $\mathbf{O}$ is random, the correlations
among them are negligible. On the other hand, when $\epsilon$ and
$1-\alpha$ are proportional to $1/N$, the number of unsuppressed modes
is of order $N$ and hence the combined effect of their correlations
become important.  Breakdown of the AA occurs when memory of early
times begins to decrease due to strong long-time interference.  Our
simulations indicate that this occurs for $\rho \equiv N(1-\alpha )
\lesssim 10$, as seen for small $\rho$ in Figs.~\ref{fig:examples} 
and \ref{fig:extensive}. Derivation of a full quenched theory requires 
appropriate handling of the intricate correlations among high powers 
of random orthogonal matrices.

\textbf{Robustness:} Our results show that systems with the DSR or
orthogonal architectures are tolerant to stochastic noise in their
network dynamics up to noise amplitudes significantly larger than
$1/N$.  An important issue is the sensitivity of the network to
structural noise. We have tested numerically the robustness of the
orthogonal recurrent network to neuron deletion.  We find that this 
perturbation does not affect drastically the capacity of the system 
provided that the output weights are retrained after the neurons' removal, 
in contrast to the simple delay line. If the output weights are not
retrained, however, capacity drops substantially.  Therefore for this
to be a viable model of working-memory, the system would need to
relearn weights $\mathbf{u}_{k}$ sufficiently quickly upon neuron loss
\cite{jaeger}. Note also that these results imply that $\mathbf{W}$
need not be exactly orthogonal since neuron removal is also a
perturbation away from orthogonality.

\begin{figure}[htb]
\centering \includegraphics[width=2.5in]{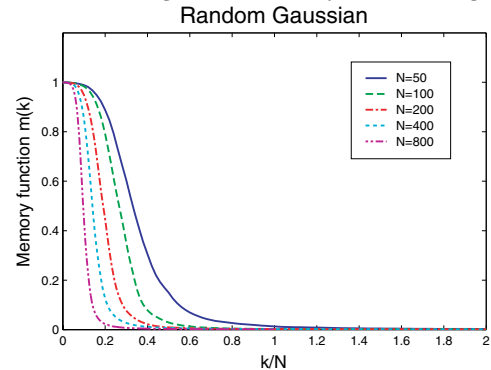}
\caption{Numerical calculation of the memory function of Gaussian random
matrices for $\bar{\protect\epsilon}=0.01$, $\protect\alpha=0.999$ and
different sizes. Results are averages over 50 realizations of the
connectivity matrix. Exploring different values of $\alpha$ we find
that up to the above mentioned value the memory improves with increasing 
$\alpha$. Increasing $\alpha$ beyond this value results in
irregular and highly variable $m(k)$.}
\label{fig:gaussian}
\end{figure}

\textbf{Random Gaussian Matrices:} In this work we have assumed rather
special network architectures. On the other hand, there are claims
that any generic (stable) connection matrix $\mathbf{W}$ can robustly
store long temporal signals \cite{maass, jaeger}; if substantiated
this is indeed a powerful result. The theoretical study of more
generic ensembles is difficult.  However, our simulations of fully
connected Gaussian random matrices indicate that their capacity is not
extensive.  If $\epsilon =0$ and $\alpha$ is sufficiently small then
it is likely that $m(k)=1$ for $k\leqslant N$ and zero for larger $N$,
as is the case in the models studied here. This is because for small
$\alpha$, interference from long past times is negligible and hence
the sum-rule Eq.~\ref{sumrule} implies the above square form for
$m(k)$.  However, this capacity is unusable in large systems because
it requires exponentially large output weights (reflecting the near
singularity of the correlation matrix $\mathbf{C}$).  Taking $\alpha$
close to 1 so that $\mathbf{C}$ is well conditioned, results in strong
fluctuations in $m(k)$ and does not seem to yield extensive capacity.
The presence of noise also regularizes the system but again does not
contribute to give extensive capacity, as indicated in
Fig.~\ref{fig:gaussian}. We are currently developing a fuller
understanding of the short-term memory properties of generic
connectivity matrices.

\textbf{Acknowledgments:} HS is partially supported by the
Israel Science Foundation (Center of Excellence 8006/00), and DDL
by the U.S. Army Research Office.  OW acknowledges
support from the National Science Foundation via DMR 02209243 and the
National Institute of Health via RO1 EY01473701. We acknowledge
helpful discussions with Misha Tsodyks and are grateful to Daniel Fisher for
many illuminating discussions on this work. We also thank our referees for 
helpful comments.

%\noindent{Acknowledgements}

\end{document}